\documentclass[aps,twocolumn,floats,prd,nofootinbib,superscriptaddress,10pt,longbibliography]{revtex4-1}
\usepackage[utf8]{inputenc}
\usepackage{graphicx}
\usepackage{dcolumn}
\usepackage{bm}
\usepackage{hyperref}
\usepackage{lipsum}
\usepackage{xcolor}
\usepackage{calc}
\usepackage{accents}
\usepackage{comment}
\usepackage{float}
\usepackage{diagbox}
\usepackage{soul}
\usepackage{ulem}

\makeatletter
\newcommand\footnoteref[1]{\protected@xdef\@thefnmark{\ref{#1}}\@footnotemark}
\makeatother

\newcommand{\Planck}{{\it Planck}}

\begin{document}

\preprint{APS/123-QED}

\title{The Sigma-8 Tension is a Drag}
\author{Vivian Poulin}
\affiliation{Laboratoire Univers \& Particules de Montpellier (LUPM), CNRS \& Universit\'e de Montpellier (UMR-5299),Place Eug\`ene Bataillon, F-34095 Montpellier Cedex 05, France}
\author{Jos\'e Luis Bernal}
\affiliation{William H.\ Miller III Department of Physics, Johns Hopkins University, 3400 North Charles Street, Baltimore, MD 21218 USA}
\author{Ely D. Kovetz}
\affiliation{Physics Department, Ben-Gurion University of the Negev, Beersheba, Israel}
\author{Marc Kamionkowski}
\affiliation{William H.\ Miller III Department of Physics, Johns Hopkins University, 3400 North Charles Street, Baltimore, MD 21218 USA}

\begin{abstract}
Measurements of weak gravitational lensing at low redshifts ($z\lesssim 0.5-1$), quantified by the parameter $S_8$, favor weaker matter clustering than that expected from the standard $\Lambda$CDM cosmological model with parameters determined by cosmic microwave background (CMB) measurements. However, the amplitude of matter clustering at higher redshifts, as probed by lensing of the CMB, is consistent with $\Lambda$CDM.  In the literature, it has been found that the tension can be resolved by introducing a friction between dark matter and dark energy without altering the tightly constrained expansion history.  Here we show that in order to get a low $S_8$ value consistent with the findings of cosmic shear under $\Lambda$CDM, cosmological measurements favor (at $\sim3\sigma$, in this one parameter model) a non-zero drag leading to a suppression of low-redshift power right precisely around the transition from matter to dark-energy domination. Our results hint at a connection between the $S_8$ tension and the long-standing `cosmic coincidence problem'.
We suggest ways to further probe the scenario.

\end{abstract}

\maketitle
\section{Introduction}
The concordance cosmological model, $\Lambda$CDM, provides an excellent description of increasingly precise measurements of the expansion history of the Universe and fluctuations in the matter density from observations of the cosmic microwave background (CMB) and the galaxy distribution.  
However, the increasing precision of observations and analyses has revealed tensions between the values of different parameters among different observables and experiments (see e.g.\ Ref.~\cite{Abdalla:2022yfr} for a recent review). 

Here we focus on the tension related with the amplitude of the matter clustering in the late Universe, parameterized with the combination $S_8\equiv \sigma_8(\Omega_{\rm m}/0.3)^{0.5}$, where $\sigma_8$ is the root mean square of the amplitude of matter perturbations smoothed over $8\,h^{-1}$Mpc, $h$ is the Hubble constant in units of $100\,{\rm  km~s}^{-1}$~Mpc$^{-1}$, and $\Omega_{\rm m}$ is the matter density parameter today. Simply stated, the distribution of galaxies and matter in the late Universe (redshifts $z\lesssim0.5-1$) as measured by low-redshift probes is smoother than expected from the evolution of the fluctuations observed in the CMB. 
These low-redshift probes include galaxy clustering\footnote{See Ref.~\cite{Simon:2022lde} for a discussion about the role of prior volume effect on the result.}~\cite{Philcox:2021kcw,Zhang:2021yna,Yuan:2022jqf, Zhai:2022yyk, Simon:2022lde}, galaxy weak lensing~\cite{HSC:2018mrq,DES:2021wwk,DES:2021bvc, DES:2021vln, Busch:2022pcx}, galaxy clusters~\cite{Planck:2015lwi}, CMB lensing tomography~\cite{Krolewski:2021yqy, White:2021yvw, Chen:2022jzq, DES:2022urg}
and their  combination and cross-correlations~\cite{Heymans:2020gsg, DES:2021zxv, Garcia-Garcia:2021unp, LSSTDarkEnergyScience:2022amt}; the derived values of $S_8$ systematically fall $\sim\! 2-3\sigma$  
lower than the values obtained from the primary CMB anisotropies~\cite{Planck:2018vyg, Aiola:2020azj} and from the power spectrum of CMB lensing alone~\cite{Planck:2018lbu,Wu:2019hek}.\footnote{Note that CMB lensing tomography involves the cross-correlation of the CMB lensing and a low-redshift tracer of the large-scale structure, usually galaxy clustering. This cross correlation in practice limits the support of the CMB lensing kernel to the redshift coverage of the galaxy sample employed.} Furthermore, the deviation of $\sigma_8$ as measured from CMB lensing tomography from the $\Lambda$CDM prediction grows as the redshift decreases~\cite{Chen:2022jzq}.

There are two features in the measurements that highly restrict potential solutions to this tension. First, the expansion history of the Universe is tightly constrained to follow the predictions of $\Lambda$CDM~\cite{Planck:2018vyg,Brout:2022vxf,Bernal:2016gxb,Poulin:2018zxs,Bernal:2021yli}, so that background quantities such as $\Omega_{\rm m}$ cannot be modified to resolve the tension (see Refs.~\cite{Colgain:2022rxy,Colgain:2022nlb} for an alternative view). Rather, a solution must decrease the clustering amplitude $\sigma_8$ \cite{Heymans:2020gsg,Schoneberg:2021qvd, Amon:2022azi}. Secondly,
the auto power spectrum of the CMB lensing, the support kernel of which spans in redshift from today to recombination (with a very wide peak at $z\sim 1-2$),
favors high values of $S_8$, compatible with those from the primary CMB anisotropies. This is also supported by eBOSS quasars data at high redshift $z\sim 1.5$, that is statistically compatible with \Planck\ \cite{Neveux:2020voa,Simon:2022csv}.

These findings indicate that any physics beyond $\Lambda$CDM that may reconcile these measurements must be limited to the perturbation level, leaving the background evolution untouched, and must be effective\footnote{ Another avenue invokes modification restricted to small-scales ($k\gtrsim0.1$h/Mpc) where current CMB data lose support, that may be active already at early-times, e.g. \cite{Joseph:2022jsf,He:2023dbn,Rogers:2023ezo}. These models could be probed by Lyman-alpha data \cite{Armengaud:2017nkf,Irsic:2017yje} or more accurate CMB lensing measurements at small scales \cite{Hlozek:2017zzf}. Finally, we mention that it has been suggested that bayonic effects may be responsible for the low-$S_8$ estimates in cosmic shear data \cite{Amon:2022azi,Arico:2023ocu}.} at $z\lesssim 1$. Coincidentally, this period corresponds to the epoch at which dark energy (DE) starts to dominate over dark matter (DM) in the total energy content of the universe. This raises the question of whether the appearance of a new phenomenon at low-$z$ that leads to low-$\sigma_8$ measurement could be tied to the beginning of DE domination, in turn providing us with new insight on the famous ``cosmic coincidence'' problem (see e.g. Refs.~\cite{Griest:2002cu,Dodelson:2001fq,Kamionkowski:2014zda}). 

In the past, it has been shown that the $\sigma_8$ tension can be resolved by a drag force between DM and DE that becomes operational at low redshifts \cite{Simpson:2010vh,Skordis:2015yra,Baldi:2016zom,Kumar:2017bpv,Asghari:2019qld,BeltranJimenez:2020qdu,Figueruelo:2021elm,BeltranJimenez:2021wbq}.  The drag slows the falling of DM into gravitational potential wells and thereby suppresses the growth of power. Here we show that in order to obtain a low $S_8$ value, the effect must occur only at low redshifts, precisely when DE becomes dynamically important.  This requires, of course, for there to be a preferred frame for the DE, and thus that the equation-of-state parameter be $w\neq -1$.  Otherwise, the model leaves the expansion history unchanged (and thus differs from ideas  \cite{Wang:2016lxa,DiValentino:2019jae,Lucca:2020fgp,Lucca:2021dxo,Nunes:2022bhn} on DE-DM interactions that affect the expansion history).   Our work differs from prior works \cite{Simpson:2010vh,Skordis:2015yra,Baldi:2016zom,Kumar:2017bpv,Asghari:2019qld,BeltranJimenez:2020qdu,Figueruelo:2021elm,BeltranJimenez:2021wbq} (which provide microphysical models for the drag) in the articulation 
 of the crucial role played by the CMB and galaxy weak lensing data, and in the use of powerful new data sets to constrain a minimal yet realistic model.

 \section{A coincidence of time and amplitude}
\subsection{model of DM and DE drag}

\begin{figure}[h!]
\centering
\includegraphics[width=1\columnwidth]{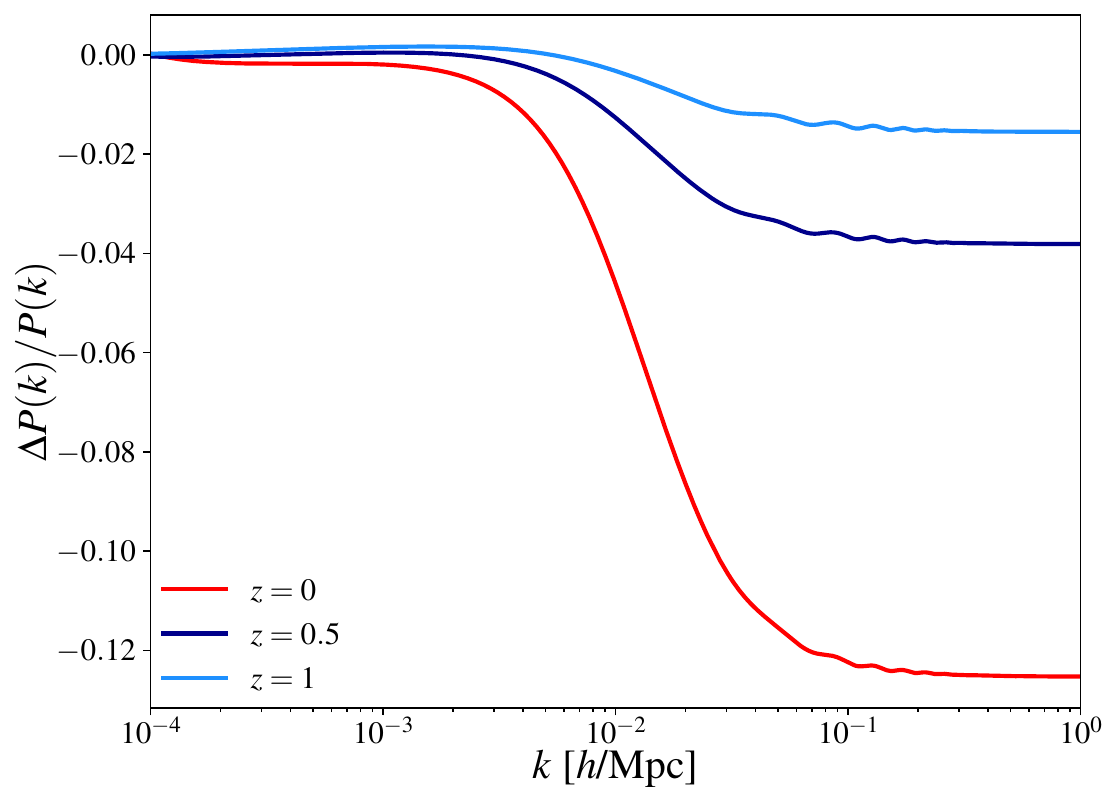}
\caption{Prediction for the power suppression in the best-fit $w$DMDE model to the full dataset including $S_8$ compared to $\Lambda$CDM, $(P(k)_{\rm DMDE}-P(k)_{\Lambda{\rm CDM}})/P(k)_{\Lambda{\rm CDM}}$ at $z=0,0.5,1$.}
\label{fig:pk}
\end{figure}

We begin by introducing the phenomenological model, similar to Refs.~\cite{Simpson:2010vh,Asghari:2019qld}. 
Working in Newtonian gauge, and following the notation of Ref.~\cite{Ma:1995ey}, we include the drag between DM and DE modifying the evolution equations for the velocity divergences $\theta$ as
\begin{eqnarray}
    \theta_{\rm DM}' &  =  & -\frac{a'}{a}\theta_{\rm DM}+k^2\psi +\Gamma_{\rm DMDE}(a)(\theta_{\rm DE}-\theta_{\rm DM}),\nonumber\\
    \theta_{\rm DE}'  &  =  &   -(1-3c_{s,{\rm DE}}^2)\frac{a'}{a}\theta_{\rm DE} +\frac{k^2c_{s,{\rm DE}}^2}{(1+w_{\rm DE})}\delta_{\rm DE}\nonumber\\
   & &
    +k^2\psi - \Gamma_{\rm DMDE}(a)R(\theta_{\rm DE}-\theta_{\rm DM}),
\end{eqnarray}
where $a$ is the scale factor, $\psi$ is the gravitational potential, $'$ denotes a derivative with respect to conformal time, and $c_{s,{\rm DE}}\equiv1$ and $w_{\rm DE}$ are the DE sound speed and equation-of-state parameter. 
We parametrize $\Gamma_{\rm DMDE}(a)$ and $R$ as~\cite{Simpson:2010vh,Asghari:2019qld}
\begin{equation}
    \Gamma_{\rm DMDE}(a)=\frac{a\Gamma_{\rm DMDE}}{\bar\rho_{\rm DM}(a)},~R=\frac{\bar\rho_{\rm DM}(a)}{(1+w_{\rm DE})\bar\rho_{\rm DE}(a)}\,,
\end{equation}
with $\bar{\rho}_i$  the mean proper energy densities of DM and DE. The scaling of the interaction rate follows from assuming a time-independent coupling constant between the 4-velocity of DM and DE at the level of the equation of motions \cite{Asghari:2019qld}. To gain some insight on the impact of this interaction, we illustrate the effect of the drag term on the matter power spectrum in Fig. \ref{fig:pk} (setting the parameters to the best-fit values extracted from our analyses with\footnote{We find that the scales below which the suppression occurs are rather sensitive to the value of $w_{\rm DE}$ due to the divergence in the equation for the DE bulk velocity. However, the amplitude of the suppression, which is what matters most when computing $S_8$, is largely unaffected by the exact value of $w_{\rm DE}$.} free $w_{\rm DE}$). The drag between DM and DE suppresses the matter power spectrum on scales that are within the horizon once the interaction becomes sizable. The suppression with respect to $\Lambda$CDM grows with time. 

The most general setup involves a free DE equation of state parameter $w_{\rm DE}$, so $w_{\rm DE}$ and $\Gamma_{\rm DMDE}$ are the additional parameters of this model. However, type-Ia supernovae (SNeIa) and baryon acoustic oscillations (BAOs) constrain $w_{\rm DE}\simeq -1$ \cite{Planck:2018vyg,Brout:2022vxf,Simon:2022csv}. We therefore start by considering a fiducial scenario where we set $w_{\rm DE}=-0.98$ (i.e. satisfying current constraints), leaving $\Gamma_{\rm DMDE}$ as our only extra parameter with respect to $\Lambda$CDM. We will later show that leaving $w_{\rm DE}$ free only has a minor impact on our conclusions.
 \subsection{Analytical argument for a new coincidence}

The drag rate  in the evolution equation of $\theta'_{\rm DM}$ becomes non-negligible when $\Gamma_{\rm DMDE}(a)/{\cal H}(a)\sim 1$, with ${\cal H}(a) \equiv a'/a$. 
We re-express the ratio between these rates as~\cite{Asghari:2019qld}
\begin{equation}
    \frac{\Gamma_{\rm DMDE}(a)}{\cal H}=\frac{\Gamma_{\rm DMDE}/(H_0\rho_c)}{E(a)\Omega_{\rm m} a^{-3}}\,,\label{eq:ratio}
\end{equation}
where  $\Omega_{\rm m} = \rho_{\rm m} / \rho_c$, with $\rho_c=(3H_0^2/8\pi G)$ the critical density, and $E(a)\equiv H(a)/H_0$. 
The amplitude of the drag rate $\Gamma_{\rm DMDE}$ is {\it a priori} unconstrained. Yet, a remarkable implication of this equation, is that for  $\Gamma_{\rm DMDE}\!\sim \!H_0\rho_c$, the drag becomes effective for\footnote{Here we approximate $\Omega_m\sim \Omega_{\rm cdm}$ for simplicity, given that we are working at the order of magnitude level.} $E(a)\Omega_{\rm m} a^{-3}\!=\!(\sqrt{\Omega_{\rm m}a^{-3}+\Omega_{\rm DE}(a)})\Omega_{\rm m}a^{-3}\! \sim\! 1$, where $\Omega_{\rm DE}$ is the DE density parameter. Assuming $w_{\rm DE}\sim -1$ and $\Omega_{\rm DE}=1-\Omega_{\rm m}\sim 0.7$, this is fulfilled specifically around $a_{\Lambda}\sim(\Omega_{\rm DE}/\Omega_{\rm m})^{-1/3}$. 
In other words, for this simple scaling of the  amplitude of the momentum drag rate, the interaction and the effects of DE become relevant around the same time, indicating that the cosmic-coincidence problem may be connected to the low values of $S_8$ measured at low-$z$. As shown below, current observations indeed favor values of $\Gamma_{\rm DMDE}$ falling in this range.
\section{Monte Carlo Analysis}
\subsection{Analysis setup}
To evaluate the success of the model under study, we perform a series of Markov-chain Monte Carlo (MCMC) runs, using the public code {\sf MontePython-v3}\footnote{\url{https://github.com/brinckmann/montepython_public}}~\citep{Audren:2012wb,Brinckmann:2018cvx}, which we interface with our modified version of {\sf CLASS}\footnote{\url{https://lesgourg.github.io/class_public/class.html}}~\cite{Lesgourgues:2011re,Blas:2011rf}. We use the Metropolis-Hasting algorithm assuming  flat priors on $\{\omega_b,\omega_{\rm cdm},100\theta_s,\log(10^{10}A_s),n_s,\tau_{\rm reio}\}+\Gamma_{\rm DMDE}$.  To test the convergence of the MCMC chains, we use the Gelman-Rubin \citep{Gelman:1992zz} criterion $|R -1|\!\lesssim\!0.01$. To post-process the chains and plot figures we use {\sf GetDist} \cite{Lewis:2019xzd}.
 
We adopt the \textit{Planck} collaboration convention in modeling free-streaming neutrinos as two massless species and one massive with $m_\nu=0.06$ eV \cite{Ade:2018sbj}. We do not include {\sf Halofit}~\cite{Smith:2002dz, Takahashi:2012em, Mead:2020vgs} to estimate the non-linear matter clustering (which is critical to model galaxy-galaxy weak lensing correlation functions), as the presence of the drag term could affect non-linear clustering. Accounting for the effects of the drag force on the non-linear matter power spectrum is beyond the scope of this paper; therefore, to minimize the impact of this limitation we consider (mostly) linear observables in our analysis.

We make use of the full \textit{Planck} 2018 TT,TE,EE and lensing power spectra \cite{Planck:2018vyg}, BAO and $f\sigma_8$ (where $f$ is the linear growth rate) measurements from BOSS DR12 \& 16 at ${z = 0.38, 0.51, 0.68}$~\cite{Alam:2016hwk,Bautista:2020ahg,Gil-Marin:2020bct,eBOSS:2020yzd}, SDSS DR7 at $z = 0.15$~\cite{Ross:2014qpa}, and QSO measurements at $z=1.48$ \cite{Neveux:2020voa,Hou:2020rse}, as well as BAO-only measurements from 6dFGS at $z = 0.106$~\cite{Beutler:2011hx} and Ly-$\alpha$ auto-correalation and cross-correlation with QSO at $z=2.334$ \cite{duMasdesBourboux:2020pck}.
We also include  uncalibrated luminosity distances to SNeIa from Pantheon+ in the range ${0.01<z<2.3}$~\cite{Brout:2022vxf}. We refer to this compilation of measurements as our fiducial data set, denoted as ${\cal D}_{\rm base}$. We also consider cases including Gaussian priors on $S_8$ as measured by KiDS-1000x\{2dFLenS+BOSS\} ($S_8=0.766^{+0.02}_{-0.014}$) ~\cite{Heymans:2020gsg} and DES-Y3 ($S_8=0.776\pm0.017$)~\cite{DES:2022urg}. 
In future work, our results shall be confirmed including the modeling of the non-linear matter power spectrum to consider the full galaxy weak lensing measurements in the analysis.  

\subsection{Results with $w$ fixed}

\begin{figure}
\centering
\includegraphics[width=1\columnwidth]{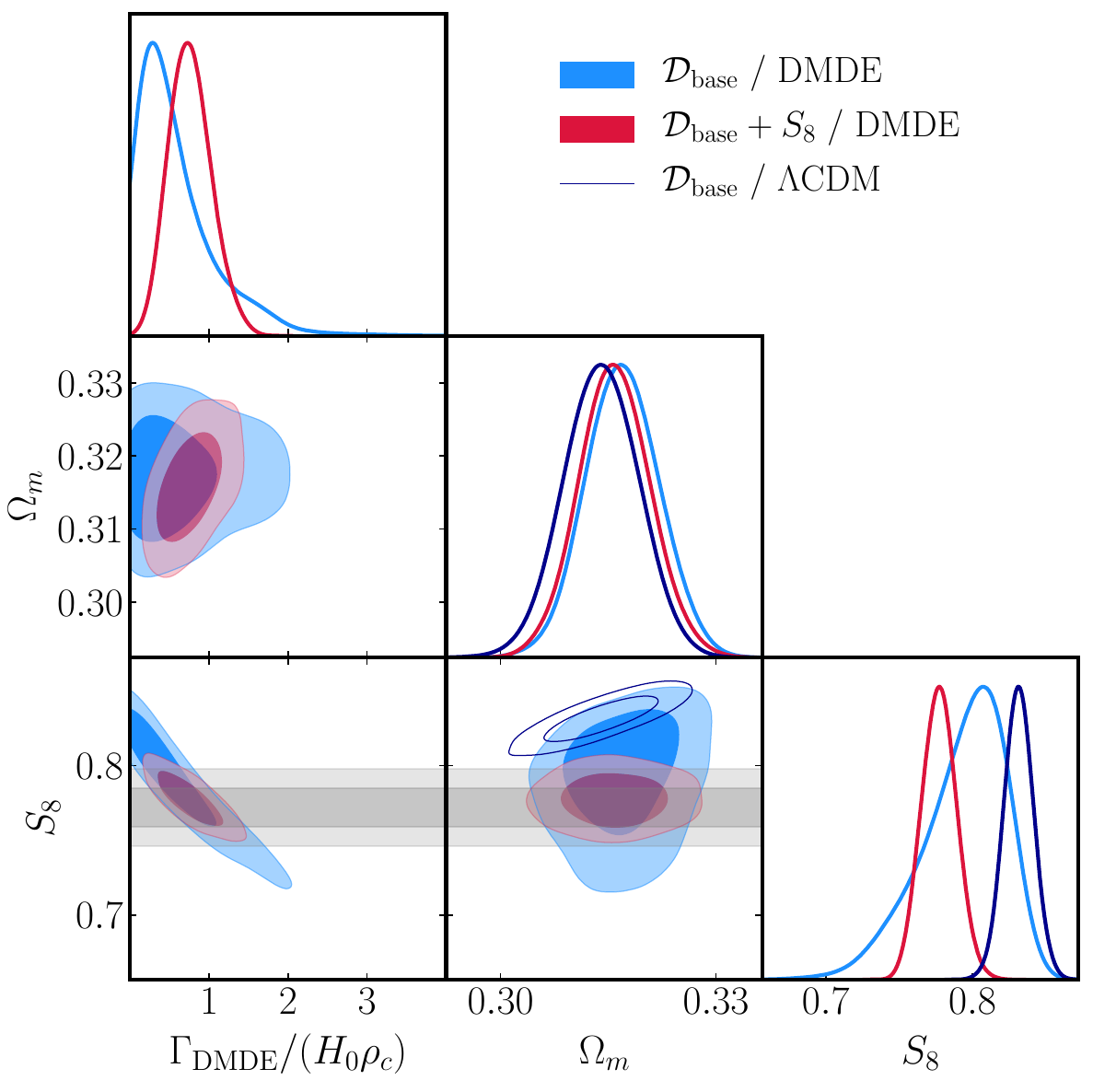}
\caption{68\% and 95\% confidence level marginalized posterior distributions of the relevant cosmological parameters for $\Lambda$CDM (dark blue lines) and the model including the drag between DM and DE (filled contours), compared with the $S_8$ measurement from galaxy lensing (grey bands) averaged assuming two independent normal distributions.}
\label{fig:MCMC}
\end{figure}

The results of our analyses with $w_{\rm DE}=-0.98$ are shown in Fig.~\ref{fig:MCMC}. We show 68\% and 95\% confidence level marginalized posterior distributions of the relevant parameters for $\Lambda$CDM and the model including the drag between DM and DE; other parameters are unchanged with respect to the standard results assuming $\Lambda$CDM. As expected, $\Gamma_{\rm DMDE}$ is very degenerate with $\sigma_8$, which allows $S_8$ to reach values even lower than the measurements of low-$z$ probes, while keeping $\Omega_{\rm m}$ effectively fixed: in particular, we find a marginalized constraint of $S_8=0.745_{-0.045}^{+0.074}$ at 68\% confidence level  for our fiducial data set. 

\begin{table*}
\def\arraystretch{1}
    \scalebox{0.9}{
    \begin{tabular}{|l|c|c|c|c|c|} 
        \hline
        Model & $\Lambda$CDM & DMDE & DMDE w/ $S_8$ & $w$DMDE & $w$DMDE w/ $S_8$ \\
        \hline
        \hline
        $\Gamma_{\rm DMDE}/(H_0\rho_c)$ & $-$ & $<1.5\, (0.0)$  &  $0.75\,(0.73)_{-0.29}^{+0.25}$ & $<2.06(0.01)$ & $0.82(1.11)_{-0.36}^{+0.27}$ \\
        $w_{\rm DE}$ & $-1$ & $-0.98$ & $-0.98$ &$<-0.95 (-0.9999)$ & $<-0.95 (-0.9999)$\\ 
        $S_8$ & $0.831\,(0.831)\pm0.011$ &    $0.796(0.830)_{-0.018}^{+0.035}$ & $0.777\,(0.778)_{-0.013}^{+0.012}$  & $0.794(0.829)_{-0.017}^{+0.039}$ & $0.776(0.776)_{-0.013}^{+0.012}$\\
        $\Omega_{\rm m}$&  $0.3139\,(0.3141)\pm0.0055$ & $0.3170\,(0.3170)_{-0.0056}^{+0.0051}$&  $0.3161\,(0.3162)_{-0.0053}^{+0.0051}$ & $0.3170(0.3145)_{-0.0062}^{+0.0056}$ &$0.3160(0.3130)_{-0.0059}^{+0.0056}$ \\
        \hline
        $\Delta \chi^2_{\rm min}$(DMDE$-\Lambda$CDM)
        & $-$  &  $+0.5$ & $-9.5$ & $0$ & $-11.5$ \\
        \hline
    \end{tabular} }
    \caption{Mean (best-fit) and $\pm$ 68\% confidence level uncertainties of the marginalized cosmological parameters for $\Lambda$CDM and DMDE (with $w_{\rm DE}=-0.98$) using our fiducial data set, and including the prior on $S_8$ in the last column. The last row shows the $\chi^2$ difference with respect to $\Lambda$CDM. }
    \label{tab:full}
\end{table*}

In Table~\ref{tab:full}, we show the mean and best-fit values of the parameters for each case, as well as the $\chi^2$ statistics. While the results show  no preference for the DM-DE drag force scenario with respect to $\Lambda$CDM for our fiducial data set (both models present similar $\chi^2$ values\footnote{The $+0.5$ degradation is coming from setting $w_{\rm DMDE}=-0.98$ in the DMDE case.}), we find $\Delta\chi^2=-9.5$ once we include the prior on $S_8$,\footnote{We find a similar $\Delta\chi^2$ for the case in which $w_{\rm DE}$ also varies.} with no statistically significant degradation in other likelihoods.\footnote{Compared to analyses without the $S_8$ prior, we find increases in $\chi^2$ of +1.9 for \Planck TTTEEE+lowl data and +0.6 for BAO+$f\sigma_8$ measurements from BOSS (which, since $\Omega_{\rm m}$ does not vary, are due to small deviations in $f\sigma_8$), that are not statistically significant.}

\subsection{Results with $w$ free}

\begin{figure}
\centering
\includegraphics[width=1\columnwidth]{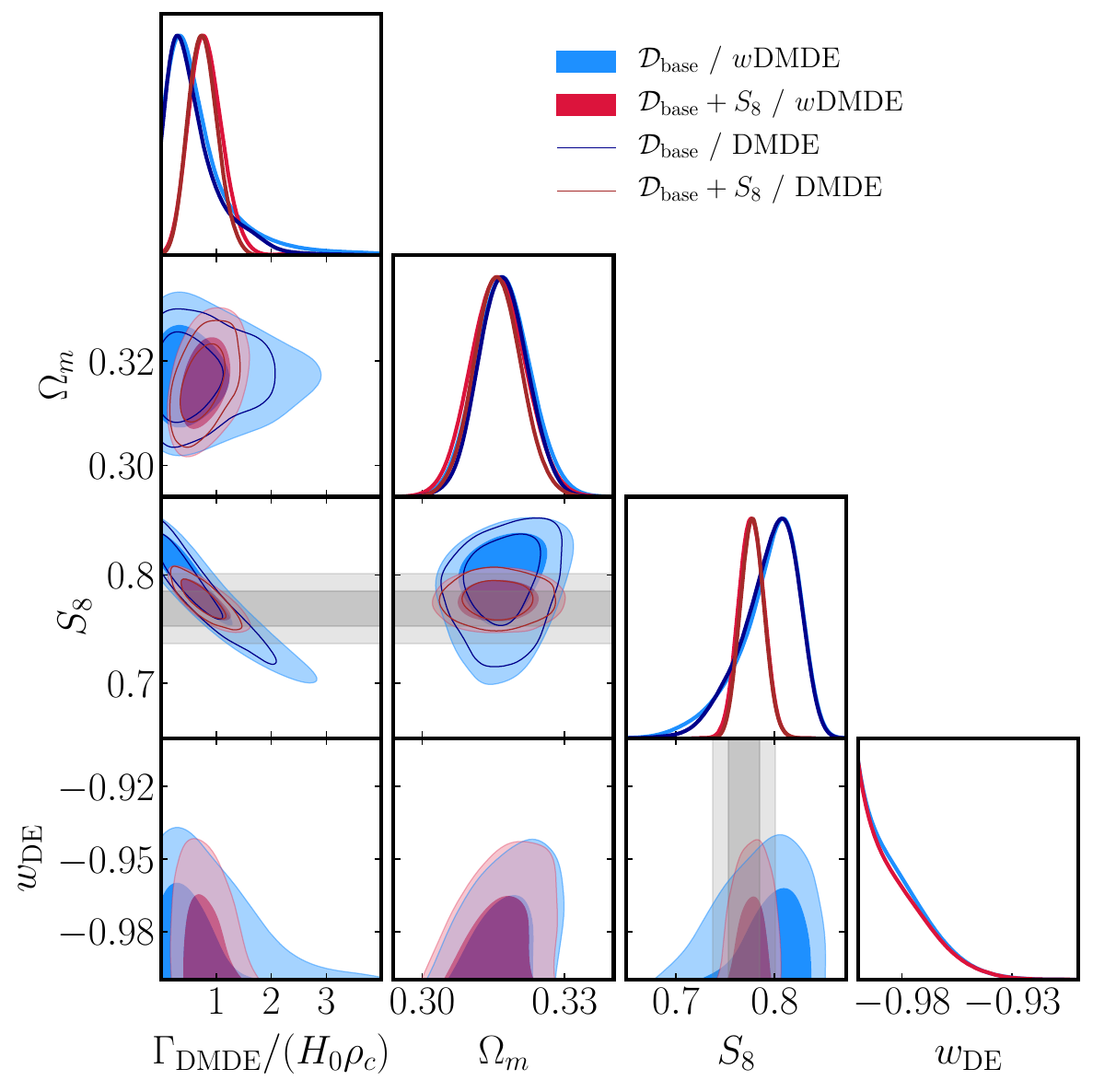}
\caption{Same as Fig.~\ref{fig:MCMC}, but now comparing results of analyses with and without $w_{\rm DE}$ let free to vary.}
\label{fig:MCMC_wfree}
\end{figure}

We now turn to the case where $w_{\rm DE}$ is let free to vary, as in full generality it is also a free parameter of the model. As previously, we report in Table~\ref{tab:full} the mean and best-fit values of the parameters $\{\Gamma_{\rm DMDE}/(H_0\rho_c),w_{\rm DE},S_8,\Omega_{\rm m}\}$, as well as the $\chi^2$ statistics, with and without including the $S_8$ prior.
As expected, data constrain $w_{\rm DE}<-0.95$ at 95\% C.L. (in good agreement with the literature \cite{Planck:2018vyg,Brout:2022vxf}) but the ability of the model to resolve the $S_8$ tension is left unchanged. We compare the results further on Fig.~\ref{fig:MCMC_wfree}, where we show the posterior distribution of the parameters of interest with and without letting $w_{\rm DE}$ free to vary. One can see that the impact of leaving $w_{\rm DE}$ free is minor given current strong constraints on deviations away from $-1$, and we find that the reconstructed value of $\Gamma_{\rm DMDE}$ (and $S_8$) is largely unaffected.
 
\section{Discussion: is the $\sigma_8$ tension a coincidence?}
\begin{figure}[h!]
\centering
\includegraphics[width=1\columnwidth]{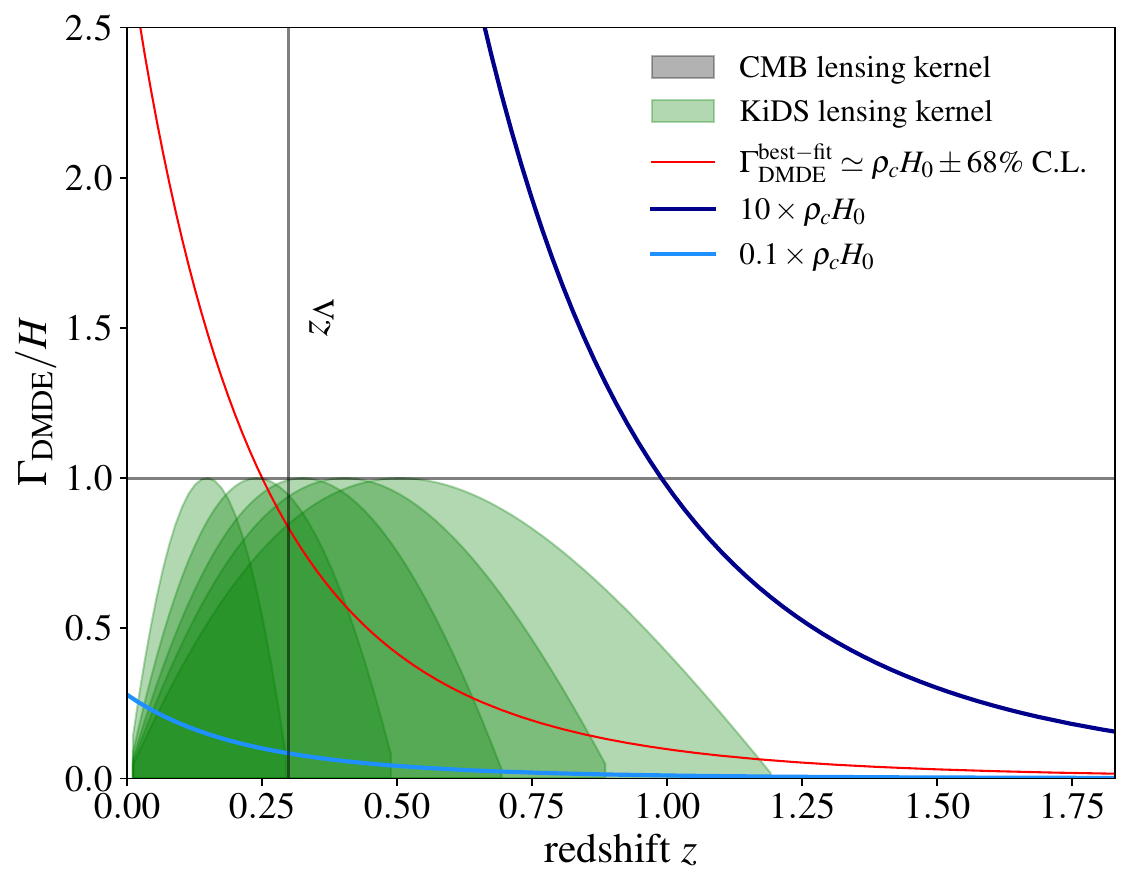}
\caption{Ratio of the interaction rate $\Gamma_{\rm DMDE}$ over the Hubble rate $H(z)$. We also show the CMB and galaxy weak lensing kernel functions (given the five $z$-bins of KiDS \cite{Hildebrandt:2020rno}) in the background. We indicate the redshift at which DE dominates, $z_{\Lambda}\equiv (\Omega_{\Lambda}/\Omega_{\rm m})^{1/3}-1$.  }
\label{fig:wf}
\end{figure}

Interestingly, irrespective of whether $w_{\rm DE}$ is kept fixed or varied, we find that $\Gamma_{\rm DMDE}$ must have values $\sim H_0\rho_c$ to reconcile $S_8$ from our fiducial data set with the results from galaxy lensing. As discussed above, this shows a potential connection between a low-$S_8$ value that is dynamically generated at low-$z$ due to a DM-DE drag and the cosmic coincidence problem. 
Let us stress again that this coincidence is highly non-trivial because the free parameter\footnote{We recall that we have shown background considerations enforce that $w_{\rm DE}$ does not play a role.} of the model $\Gamma_{\rm DMDE}$ is responsible for setting both the amplitude of the suppression (i.e. whether we can achieve the $\sigma_8$ measured by weak lensing) and the time at which $\Gamma \sim H$ (i.e. when the interaction becomes relevant), and $\Gamma_{\rm DMDE}$ is not a priori constrained.  We illustrate this in Fig.~\ref{fig:wf}: if $\Gamma_{\rm DMDE}$ were much larger, the interaction would have modified the evolution of perturbations at higher redshifts, potentially causing tension with the large-scale CMB lensing power spectrum (and even the integrated Sachs-Wolfe effect). On the other hand, a much lower $\Gamma_{\rm DMDE}$ would leave $S_8$ unaffected because the interaction would not be relevant at any time.  DM-DE drag occurring right at the onset of DE domination can therefore explain why the only data that are affected are those probing perturbations dynamics at $z < 0.5$, with the right amplitude of suppression, and not higher redshift ones which are left unaffected.
Hence, this scenario suggests that the transition to a DE-dominated epoch may have non-trivial implications for model building,  potentially guiding   future research, with some attempts already discussed in Refs.~\cite{Simpson:2010vh,Wang:2016lxa,Asghari:2019qld,BeltranJimenez:2021wbq}.

\begin{figure}[h!]
\centering
\includegraphics[width=1\columnwidth]{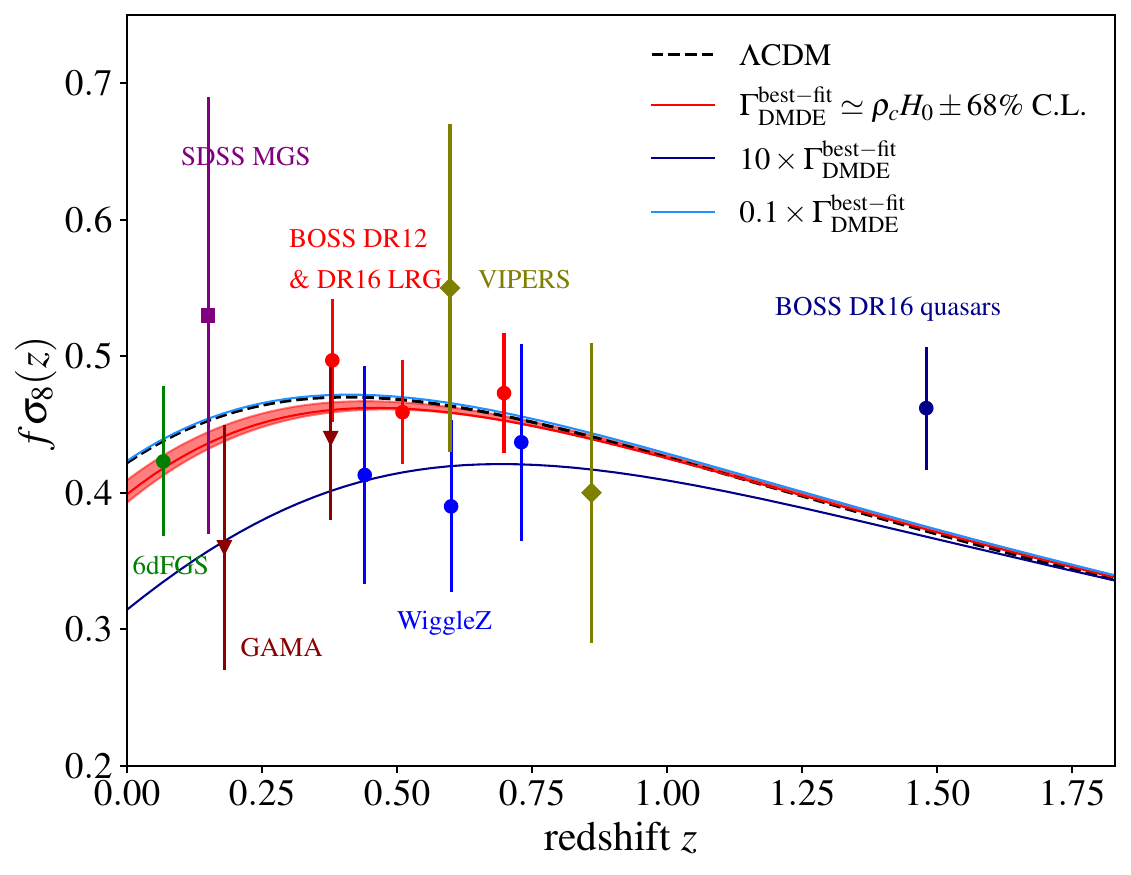}
\caption{Prediction for $f\sigma_8$ in the $\Lambda$CDM and DMDE models compared to a sample of data. Our analysis use 6dFGS, SDSS MGS, BOSS DR12 \& 16 (LRG and quasars) data.}
\label{fig:fs8}
\end{figure}

Importantly, this model can be further tested with future higher-accuracy measurements.   Forthcoming CMB and galaxy surveys may be able to discriminate between the models.  
Additionally, the redshift-dependent $\sigma_8$ deviation from the prediction of $\Lambda$CDM can also be probed with CMB lensing tomography and $f\sigma_8$ measurements. Forthcoming low-$z$ clustering measurements, e.g.\ from the DESI bright galaxy sample~\cite{DESI:2016fyo}, may detect the deviation in $f\sigma_8$ (see Fig.~\ref{fig:fs8}). 
Interestingly, the prediction for $f\sigma_8(z)$ is similar to that obtained in Ref.~\cite{Nguyen:2023fip}, in which authors studied a model where the growth index is left free to vary (and the background dynamics is identical to $\Lambda$CDM), showing that current data favor a deviation of the growth index from $\Lambda$CDM at 3.7$\sigma$.
Similarly, improved cross-correlations between large-scale-structure surveys and CMB lensing will also weigh in on this.

Looking forward, there are additional observables in which the DM-DE drag would leave signatures that could be searched for to probe this scenario, if the effects are properly modeled.
First, as mentioned above, this model shall be applied to the full set of measurements from galaxy lensing surveys, rather than just a prior on $S_8$.
This will require improved modeling of the DM-DE drag including non-linear clustering (see e.g. Refs.~\cite{Baldi:2014ica,Baldi:2016zom}). Meanwhile, similarly to the case of self-interacting DM~\cite{Tulin:2017ara}, the friction between DM and DE would affect the intrinsic alignment of galaxies. Including the DM-DE drag in the formalism of intrinsic alignments (see e.g., Ref.~\cite{Vlah:2019byq}) would allow to use correlations of galaxy shapes to probe this model, distinguishing it from the effects of baryonic physics~\cite{Harvey:2021gbi}. 
In addition, on larger scales, the suppression of the potential wells induced by the drag could lead to an interesting integrated Sachs-Wolfe signal (too small when considering the CMB TT power spectrum due to cosmic variance) but that could be picked-up when cross-correlating with CMB lensing or galaxy distributions. Finally, since baryons are unaffected by this interaction, it is expected that DM spirals down and collapses faster than baryons, leaving signatures like modified tidal streams, as well as potentially affecting probes of DM density profiles such as stellar-rotation velocities, and introducing a small velocity bias between DM and baryons that can be searched for in galaxy clustering measurements~\cite{Desjacques:2016bnm} and galaxy clusters~\cite{Anbajagane:2021gfx}. Evaluating the scope of these searches and their sensitivity to the DM-DE drag provides very motivated targets for future studies. Let us also mention the possibility of extending the model to consider a drag between DE and baryons \cite{Vagnozzi:2019kvw,Jimenez:2020ysu,Ferlito:2022mok}.

\section{Conclusions}
To conclude, we have explored a cosmological model that shows promise to resolve the tension in clustering between high-redshift and low-redshift probes. 
While further study is required, both modeling the non-linear effects in the matter power spectrum and using additional cosmological and astrophysical probes, we have bolstered the motivation for this model. 
We showed that the required values to reconcile high-redshift measurements with the $S_8$ values preferred by galaxy-lensing observations naturally set the moment in which the DM-DE interaction becomes effective to be around DM-DE equality. 
Hence, this model involves interesting phenomenology not only for the dark sector of the Universe but it may also imply that the last transition in the history of the Universe could have had non-trivial implications. 
We hope this work spurs interest in this family of solutions for the $S_8$ tension, both regarding model building and adding detail to the astrophysical and cosmological consequences of the friction to conclusively distinguish it from $\Lambda$CDM.

\begin{acknowledgements}
We thank Marco Baldi, Dario Bettoni, David Figueruelo, Jose Beltran Jimenez, Antony Lewis, Matteo Lucca, Florencia A. Teppa Pannia and Sunny Vagnozzi for useful comments on the manuscript.
The authors acknowledge the use of computational resources from the Excellence Initiative of Aix-Marseille University (A*MIDEX) of the “Investissements d’Avenir” programme. 
This project has received support from the European Union’s Horizon 2020 research and innovation program under the Marie Skodowska-Curie grant agreement No 860881-HIDDeN. This work has been partly supported by the CNRS-IN2P3 grant Dark21.  This project has received funding from the European Research Council (ERC) under the
European Union’s HORIZON-ERC-2022 (Grant agreement No. 101076865).  JLB was supported by the Allan C. and Dorothy H. Davis Fellowship. EDK was supported by a faculty fellowship from the Azrieli Foundation. 
MK was supported by NSF Grant No.\ 2112699 and the Simons Foundation.
\end{acknowledgements}

\bibliography{biblio}

\end{document}